\documentclass[draft,tightenlines,nofootinbib,preprint,aps,eqsecnum,amsmath,amssymb]{revtex4}

\newcommand{\beq}{\begin{equation}}
\newcommand{\eeq}{\end{equation}}
\newcommand{\bea}{\begin{eqnarray}}
\newcommand{\eea}{\end{eqnarray}}
\newcommand{\cir}{{\buildrel \circ \over =}}
\newcommand{\sgn}{\mbox{\boldmath $\epsilon$}}

\begin{document}

\baselineskip 18pt

\title{Massless Particles plus Matter in the Rest-Frame Instant Form of Dynamics.}

\medskip

\author{David Alba}
\affiliation{
Sezione INFN di Firenze\\Polo Scientifico, via Sansone 1\\
 50019 Sesto Fiorentino, Italy\\
 E-mail ALBA@FI.INFN.IT}

\author{Horace W. Crater}
\affiliation{The University of Tennessee Space Institute \\
Tullahoma, TN 37388 USA \\
E-mail: hcrater@utsi.edu}

\author{Luca Lusanna}
\affiliation{ Sezione INFN di Firenze\\ Polo Scientifico\\ Via Sansone 1\\
50019 Sesto Fiorentino (FI), Italy\\ Phone: 0039-055-4572334\\
FAX: 0039-055-4572364\\ E-mail: lusanna@fi.infn.it}

\begin{abstract}

After introducing the parametrized Minkowski theory describing a
positive-energy scalar massless particle, we study the rest-frame
instant form of dynamics of such a particle in presence of another
massive one (to avoid the front form of dynamics). Then we describe
massless particles with Grassmann-valued helicity and their
quantization.

\end{abstract}

\maketitle

\vfill\eject

\section{Introduction}

The problem of the description of relativistic particles with a
complete control of the Poincare' group has been recently fully
solved by means of parametrized Minkowski theories \cite{1,2,3,4}
and of the associated inertial rest-frame instant form of dynamics
\cite{5,6}, subsequently extended to non-inertial frames \cite{7}.
The basic idea is to use admissible 3+1 splittings of Minkowski
space-time to define global non-inertial frames \cite{8,9}.

\bigskip

In the 3+1 point of view \cite{9} we assign:\medskip

 a) the world-line of an arbitrary time-like observer;

 b) an admissible 3+1 splitting of Minkowski space-time, namely a
nice foliation with space-like instantaneous 3-spaces (i.e. a clock
synchronization convention) tending to the same space-like
hyper-plane at spatial infinity (so that there are always asymptotic
inertial observers to be identified with the fixed stars). See
Ref.\cite{7} for the meaning of the M$\o$ller conditions \cite{8,9}
defining the admissible foliations.\medskip

This allows one to define a {\it global non-inertial frame} centered
on the observer and to use observer-dependent Lorentz-scalar {\it
radar 4-coordinates} $\sigma^A = (\tau ;\sigma^r)$, where $\tau$ is
a monotonically increasing function of the proper time of the
observer and $\sigma^r$ are curvilinear 3-coordinates on the 3-space
$\Sigma_{\tau}$ having the observer as origin. If $x^{\mu} \mapsto
\sigma^A(x)$ is the coordinate transformation from the inertial
Cartesian 4-coordinates $x^{\mu}$ to radar coordinates, its inverse
$\sigma^A \mapsto x^{\mu} = z^{\mu}(\tau ,\sigma^r)$ defines the
embedding functions $z^{\mu}(\tau ,\sigma^r)$ describing the
instantaneous 3-spaces $\Sigma_{\tau}$ as embedded 3-manifold into
Minkowski space-time. The induced 4-metric on $\Sigma_{\tau}$ is the
following functional of the embedding $g_{AB}(\tau ,\sigma^r) =
[z^{\mu}_A\, \eta_{\mu\nu}\, z^{\nu}_B](\tau ,\sigma^r)$, where
$z^{\mu}_A = \partial\, z^{\mu}/\partial\, \sigma^A$ and
$\eta_{\mu\nu} = \sgn\, (+---)$ is the flat metric ($\sgn = \pm 1$
according to either the particle physics $\sgn = 1$ or the general
relativity $\sgn = - 1$ convention). While the 4-vectors
$z^{\mu}_r(\tau ,\sigma^u)$ are tangent to $\Sigma_{\tau}$, so that
the unit normal $l^{\mu}(\tau ,\sigma^u)$ is proportional to
$\epsilon^{\mu}{}_{\alpha \beta\gamma}\, [z^{\alpha}_1\,
z^{\beta}_2\, z^{\gamma}_3](\tau ,\sigma^u)$, we have
$z^{\mu}_{\tau}(\tau ,\sigma^r) = [N\, l^{\mu} + N^r\,
z^{\mu}_r](\tau ,\sigma^r)$ ($N(\tau ,\sigma^r) = \sgn\,
[z^{\mu}_{\tau}\, l_{\mu}](\tau ,\sigma^r)$ and $N_r(\tau ,\sigma^r)
= - \sgn\, g_{\tau r}(\tau ,\sigma^r)$ are the lapse and shift
functions).\medskip

The 4-metric $g_{AB}(\tau ,\sigma^u )$ on $\Sigma_{\tau}$ has the
components $\sgn\, g_{\tau\tau} = N^2 - h^{rs}\, N_r\, N_s$, $-
\sgn\, g_{\tau r} = N_r = h_{rs}\, N^s$, $ - \sgn\, g_{rs} =
h_{rs}$. The inverse of $h_{rs}$, whose signature is $(+++)$, is
$h^{rs}$ ($h^{ru}\, h_{us} = \delta^r_s$) and we have
$z^{\mu}_{\tau} = N\, l^{\mu} + h^{rs}\, N_s\, z^{\mu}_r$ and
$\eta^{\mu\nu} = \sgn\, \Big(l^{\mu}\, l^{\nu} - z^{\mu}_r\,
h^{rs}\, z^{\nu}_s\Big)$. The components $g_{AB}$ play the role of
the {\it inertial potentials} generating the relativistic apparent
forces in the non-inertial frame \cite{7}.\medskip

Let us now consider any isolated system (particles, strings, fields,
fluids) admitting a Lagrangian description. Then the coupling to an
external gravitational field allows the determination of the matter
energy-momentum tensor and of the ten conserved Poincare' generators
$P^{\mu}$ and $J^{\mu\nu}$ (assumed finite) of every configuration
of the system. Let us replace the external gravitational 4-metric in
the coupled Lagrangian with the 4-metric $g_{AB}(\tau ,\sigma^r)$ of
an admissible 3+1 splitting of Minkowski space-time and let us
replace the matter fields with new ones knowing the instantaneous
3-spaces $\Sigma_{\tau}$.

For instance a Klein-Gordon field $\tilde \phi (x)$ will be replaced
with $\phi(\tau ,\sigma^r) = \tilde \phi (z(\tau ,\sigma^r))$; the
same for every other field.

Instead for a relativistic particle with world-line $x^{\mu}(\tau )$
we must make a choice of its energy sign and it will be described by
3-coordinates $\eta^r(\tau )$ defined by the intersection of the
world-line with $\Sigma_{\tau}$: $x^{\mu}(\tau ) = z^{\mu}(\tau
,\eta^r(\tau ))$.\medskip

In this way we get a Lagrangian depending on the given matter and on
the embedding $z^{\mu}(\tau ,\sigma^r)$ and this formulation has
been called {\it parametrized Minkowski theories} \cite{1,8,9}.
These theories are invariant under frame-preserving diffeomorphisms,
so that there are four first-class constraints (an analogue of the
super-Hamiltonian and super-momentum constraints of canonical
gravity) implying that the embeddings $z^{\mu}(\tau ,\sigma^r)$ are
{\it gauge variables}. As a consequence, all the admissible
non-inertial frames are gauge equivalent, namely physics does {\it
not} depend on the clock synchronization convention and/or the
choice of the 3-coordinates in $\Sigma_{\tau}$: only the appearances
of phenomena change by changing the notion of instantaneous 3-space.

\bigskip

A particular case of this description is the {\it inertial
rest-frame instant form of dynamics for isolated systems}
\cite{1,8,9} which is done in the intrinsic inertial rest frame of
their configurations: these instantaneous 3-spaces, named Wigner
3-space due to the fact that the 3-vectors inside them are Wigner
spin-1 3-vectors, are orthogonal to the conserved 4-momentum of the
configuration (in Ref.\cite{7} there is the extension to
non-inertial rest frames).\medskip

In these rest frames there are only three notions of collective
variables, which can be built by using {\it only} the Poincare'
generators (they are {\it non-local} quantities knowing the whole
$\Sigma_{\tau}$) \cite{1}: the canonical non-covariant Newton-Wigner
center of mass (or center of spin), the non-canonical covariant
Fokker-Pryce center of inertia and the non-canonical non-covariant
M$\o$ller center of energy. All of them tend to the Newtonian center
of mass in the non-relativistic limit. See Ref.\cite{9} for the
M$\o$ller non-covariance world-tube around the Fokker-Pryce 4-vector
identified by these collective variables. As shown in
Refs.\cite{4,5,6} these three variables can be expressed as known
functions of the rest time $\tau$, of  canonically conjugate Jacobi
data (frozen Cauchy data) $\vec z = Mc\, {\vec x}_{NW}(0)$ (${\vec
x}_{NW}(\tau )$ is the standard Newton-Wigner 3-position) and $\vec
h = \vec P/Mc$, of the invariant mass $Mc = \sqrt{\sgn\, P^2}$ of
the system and of its rest spin ${\vec {\bar S}}$.

As a consequence, every isolated system (i.e. a closed universe) can
be visualized as a decoupled non-covariant collective (non-local and
therefore un-observable) pseudo-particle described by the frozen
Jacobi data $\vec z$, $\vec h$ carrying a {\it pole-dipole
structure}, namely the invariant mass and the rest spin of the
system, and with an associated external realization  of the
Poincare' group.\medskip

The universal breaking of Lorentz covariance is connected to this
decoupled non-local collective variable and is irrelevant because
all the dynamics of the isolated system lives inside the Wigner
3-spaces and is Wigner-covariant. Inside these Wigner 3-spaces the
system is described by an internal 3-center of mass with a conjugate
3-momentum  and by relative variables and there is an unfaithful
internal realization of the Poincare' group \cite{5}: the internal
3-momentum, conjugate to the internal 3-center of mass, vanishes due
the rest-frame condition. To avoid a double counting of the center
of mass, i.e. an external one and an internal one, also the internal
(interaction-dependent) internal Lorentz boosts vanish. As shown in
Ref.\cite{5} the only non-zero internal generators are the invariant
mass and the rest spin and the dynamics is re-expressed only in
terms of internal Wigner-covariant relative variables. Moreover this
construction implies that the time-like observer, origin of the
3-coordinates on the Wigner 3-spaces, must be identified with the
Fokker-Pryce inertial observer \cite{5}, so that the embedding
describing the Wigner 3-spaces is $z^{\mu}_W(\tau, \sigma^u) =
Y^{\mu}(0) + h^{\mu}\, \tau + \epsilon^{\mu}_r(\vec h)\, \sigma^r$,
$h^{\mu} = (\sqrt{1 + {\vec h}^2}; \vec h)$, $\epsilon^{\mu}_r(\vec
h) = \Big(- h_r; \delta^i_r - {{h^i\, h_r}\over {1 + \sqrt{1 + {\vec
h}^2}}}\Big)$, $Y^{\mu}(0) = \Big(\sqrt{1 + {\vec h}^2}\, {{\vec h
\cdot \vec z}\over {Mc}}; {{\vec z}\over {Mc}} + {{\vec h \cdot \vec
z}\over {Mc}}\, \vec h + {{{\vec {\bar S}} \times \vec h}\over {Mc\,
(1 + \sqrt{1 + {\vec h}^2})}}\Big)$.

In the case of relativistic particles the reconstruction of their
world-lines requires a complex interaction-dependent procedure
delineated in Ref.\cite{4}. See Ref.\cite{5} for the comparison with
the other formulations of relativistic mechanics developed for the
study of the problem of {\it relativistic bound states}.

\bigskip

This allows one to get a relativistic formulation of atomic physics
as an effective theory below the threshold of pair production.  As a
consequence it is possible to include the "light-cone" in atomic
physics. While this is irrelevant for experiments on Earth, it is
fundamental for space physics in the Solar system. The ACES mission
of ESA will make the first precision measurement of the
gravitational redshift near Earth (deviation of the null geodesics
for light rays from the Minkowski ones at the $1/c^2$ order) by
putting the Pharao atomic clock on the Space Station \cite{10}.
Protocols for teleportation from Earth to the Space Station
\cite{11} will require "photon" (eikonal approximation to light
rays) propagating along null geodesics and a relativistic
formulation of entanglement \cite{12}. Finally this framework will
also be needed for the search of gravitational waves by means of
atom interferometers \cite{13}.
\bigskip

What is still missing is the description of massless particles in
the rest-frame instant form of dynamics. This will be done in this
paper.\bigskip

In Section I we describe a scalar positive-energy massless particle
by means of a parametrized Minkowski theory.\medskip

In Section II, after the addition of a decoupled positive-energy
massive particle, to avoid the front form of dynamics, we are able
to define the rest-frame instant form of the dynamics of the
isolated system of a massless particle plus a massive one.\medskip

In Section III we give the pseudo-classical description of a
massless particle carrying a Grassmann-valued helicity. Then in
Section IV we quantize the system to get a quantum particle with
helicity $\pm 1$.\medskip

In the Conclusions we make some final comments.\medskip

In Appendix A there is a review of the light-like polarization
vectors.

\vfill\eject

\section{A Scalar Positive-Energy Massless Particle}

Let the massless particle have the world-line $x^{\mu}(\tau) =
z^{\mu}(\tau, \vec \eta(\tau))$, so that it is identified by the
3-coordinates $\eta^r(\tau)$ inside the instantaneous 3-spaces
$\Sigma_{\tau}$. Usually it is describe with the singular Lagrangian
$S = \int d\tau\, {{{\dot x}^2(\tau)}\over {\mu (\tau)}}$, with $\mu
(\tau)$ Lagrange multiplier, which implies ${\dot x}^2(\tau) = 0$ as
an Euler-Lagrange equation and the first class constraint $P^2
\approx 0$ at the Hamiltonian level ($P^{\mu}$ is the 4-momentum
conjugated to $x^{\mu}(\tau)$).
\medskip

In parametrized Minkowski theories the massless particles must be
assumed of positive-energy and is described by the 3-coordinates
$\eta^r(\tau)$ on $\Sigma_{\tau}$. The previous action is now
replaced by the following one ($l^{\mu} = {1\over {det\, h_{rs}}}\,
\epsilon^{\mu}{}_{\alpha\beta\gamma}\, z^{\alpha}_1\, z^{\beta}_2\,
z^{\gamma}_3$, $l_{\mu}\, z^{\mu}_{\tau} = \sgn\, N$)

\bea
 S &=& \int d\tau\, L(\tau),\nonumber \\
 &&{}\nonumber \\
 L(\tau) &=& {1\over {\mu (\tau)}}\, {{g_{\tau\tau} + 2\, g_{\tau r}\,
 {\dot \eta}^r(\tau) + g_{rs}\, {\dot \eta}^r(\tau)\, {\dot
 \eta}^s(\tau)}\over {\sgn\, l_{\mu}\, z^{\mu}_{\tau}}}(\tau,
 \eta^u(\tau )),
 \label{2.1}
 \eea

\noindent which is a functional depending on the Lagrangian
variables $z^{\mu}(\tau, \sigma^u)$, $\eta^u(\tau)$ and $\mu(\tau)$.
\bigskip

The Euler-Lagrange equation ${{\delta\, S}\over {\delta\,
\mu(\tau)}} \, = \, - {1\over {\mu^2(\tau)\, N(\tau,
\eta^u(\tau))}}\, \Big[g_{\tau\tau} + 2\, g_{\tau r}\,
 {\dot \eta}^r(\tau) + g_{rs}\, {\dot \eta}^r(\tau)\, {\dot
 \eta}^s(\tau)\Big](\tau, \eta^u(\tau))\, \cir\, 0$ implies
 ${\dot x}^2(\tau)\, \cir\, 0$. On the Wigner 3-spaces, where
 $x^{\mu}(\tau) = Y^{\mu}(0) + h^{\mu}\, \tau + \epsilon^{\mu}_r(\vec
 h)\, \eta^r(\tau)$, this implies ${\dot {\vec \eta}}^2(\tau) \,
 \cir\, 1$.

\bigskip

The canonical momenta are

\bea
 \kappa_r(\tau) &=& {{\partial\, L(\tau)}\over {\partial\, {\dot \eta}^r}}
 = {2\over {\mu(\tau)\, N(\tau, {\vec \eta}(\tau))}}\, \Big(g_{\tau r} +
 g_{rs}\, {\dot \eta}^s(\tau)\Big)(\tau, \eta^u(\tau)) =\nonumber \\
 &=& - {{2\, \sgn}\over {\mu(\tau)\, N(\tau, {\vec \eta}(\tau))}}\,
 \Big(N_r + h_{rs}\, {\dot \eta}^s(\tau)\Big)(\tau, \eta^u(\tau)),\nonumber \\
 &&{}\nonumber \\
 \rho_{\mu}(\tau, \sigma^u) &=& {{\delta\, S}\over {\delta\,
 z^{\mu}_{\tau}(\tau, \sigma^u)}} =
  {{\delta^3(\sigma^u - \eta^u (\tau))}
 \over {\mu(\tau)\, N(\tau, \sigma^v)}}\, \Big[2\, (z_{\tau \mu} +
 z_{r \mu}\, {\dot \eta}^r(\tau)) -\nonumber \\
 &-& \sgn\,l_{\mu}\, {{g_{\tau\tau} + 2\, g_{\tau r}\,
 {\dot \eta}^r(\tau) + g_{rs}\, {\dot \eta}^r(\tau)\, {\dot
 \eta}^s(\tau)}\over N}\Big](\tau, \sigma^v),\nonumber \\
 &&{}\nonumber \\
 \pi(\tau) &=& {{\partial\, L(\tau)}\over {\partial\, {\dot \mu}}} = 0,
 \label{2.2}
 \eea

\noindent and the canonical Hamiltonian vanishes, $H_c = 0$.
\bigskip

Therefore we have the following primary constraints

\bea
 &&{\cal H}_{\mu}(\tau, \sigma^u) =
 \rho_{\mu}(\tau, \sigma^u) - \delta^3(\sigma^v - \eta^v
 (\tau))\, \Big[l_{\mu}\, \Big({1\over {\mu(\tau)}} +
 {{\mu(\tau)}\over 4}\, h^{rs}\, \kappa_r\, \kappa_s\Big)
 - \sgn\, z_{r\, \mu}\, h^{rs}\, \kappa_s\Big](\tau, \sigma^u)
 \approx 0,\nonumber \\
 &&{}\nonumber \\
 &&\pi(\tau) \approx 0.
 \label{2.3}
 \eea

\medskip

Therefore the Dirac Hamiltonian is

\beq
 H_D = \lambda(\tau)\, \pi(\tau) + \int d^3\sigma\,
 \lambda^{\mu}(\tau, \sigma^u)\, {\cal H}_{\mu}(\tau,
 \sigma^u),
 \label{2.4}
 \eeq

 \noindent where $\lambda (\tau)$ and $\lambda^{\mu}(\tau,
 \sigma^u)$ are Dirac multipliers.

\bigskip

The preservation in $\tau$ of the primary constraints ($\{ {\cal
H}_{\mu}(\tau, \sigma^u), H_D \} = 0$, $\{ \pi(\tau), H_D\} = 0$)
implies  the following secondary constraint

\beq
 \chi(\tau)\, = {1\over {\mu^2(\tau)}} - {1\over 4}\,
 h^{rs}(\tau, \eta^u(\tau))\, \kappa_r(\tau)\,
 \kappa_s(\tau) \approx 0,
 \label{2.5}
 \eeq

\noindent so that we have $\mu(\tau) \approx {2\over
{\sqrt{h^{rs}(\tau, \eta^u(\tau))\, \kappa_r(\tau)\,
 \kappa_s(\tau)}}}$.\medskip

The $\tau$-preservation of the secondary constraint $\chi(\tau)
\approx 0$ determines the Dirac multiplier $\lambda (\tau)$

\bea
 &&\lambda
(\tau) \approx \Big({1\over {2\, (h^{rs}\, \kappa_r(\tau)\,
\kappa_s(\tau))^3/2}}\, \Big[l^{\mu}\, \sqrt{h^{uv}\,
\kappa_u(\tau)\, \kappa_v(\tau)} - \sgn\, z^{\mu}_u\, h^{uv}\,
\kappa_v(\tau)\Big]\Big)(\tau, \eta^k(\tau))\, {{\partial\,
\lambda_{\mu}(\tau, \eta^k(\tau))}\over {\partial\, \eta^u}}.
 \nonumber \\
 &&{}
 \label{2.6}
 \eea

\bigskip

In conclusion the constraints ${\cal H}_{\mu}(\tau, \sigma^u)
\approx 0$ are first class. Instead the two constraints $\pi(\tau)
\approx 0$ and $\chi(\tau) \approx 0$ are second class, $\{
\chi(\tau), \pi(\tau)\} = - {2\over {\mu^3(\tau)}} \not= 0$.\medskip

If we eliminate the variables $\mu(\tau)$ and $\pi(\tau)$ by going
to Dirac brackets, the first class constraints take the following
form

\beq
 {\cal H}_{\mu}(\tau, \sigma^u) =
 \rho_{\mu}(\tau, \sigma^u) - \delta^3(\sigma^v - \eta^v
 (\tau))\, \Big[l_{\mu}\, \sqrt{h^{rs}\, \kappa_r(\tau)\, \kappa_s(\tau)}
 - \sgn\, z_{r\, \mu}\, h^{rs}\, \kappa_s(\tau)\Big](\tau, \sigma^u)
 \approx 0.
 \label{2.7}
 \eeq

\bigskip

Since, as shown in Ref.\cite{5}, the Poincare' generators generated
by the action (\ref{2.1}) are $P^{\mu} = \int d^3\sigma\,
\rho^{\mu}(\tau, \sigma^u)$ and $J^{\mu\nu} = \int d^3\sigma\,
\Big(z^{\mu}\, \rho^{\nu} - z^{\nu}\, \rho^{\mu}\Big)(\tau,
\sigma^u)$, we have that Eq.(\ref{2.7}) implies $P^{\mu} \approx
\sqrt{h^{rs}(\tau, \eta^u(\tau))\, \kappa_r\, \kappa_s}\,
l^{\mu}(\tau, \eta^u(\tau)) - \sgn\, z^{\mu}_r(\tau, \eta^u(\tau))\,
h^{rs}(\tau, \eta^u(\tau))\, \kappa_s$ and $P^2 \approx 0$. This
implies that for an isolated massless particle we cannot have the
description in the rest-frame instant form of dynamics, requiring
$\sgn\, P^2 > 0$, but only a front (null) form \cite{14} (see also
Ref.\cite{15} and its bibliography).
\medskip

However if we add massive matter it is possible to have the massless
particle described in  the rest-frame instant form of the overall
system. To this end let us add a decoupled positive-energy scalar
massive particle $x_1^{\mu}(\tau) = z^{\mu}(\tau, \eta^u_1(\tau))$
with mass $m_1$. The action (\ref{2.1}) is replaced by the following
one

\bea
 S &=& \int d\tau\, d^3\sigma\, \Big(- m_1\, c\, \delta^3(\sigma^u - \eta^u_1(\tau))\,
 \sqrt{\sgn\, [g_{\tau\tau} + 2\, g_{\tau r}\, {\dot \eta}_1^r(\tau) + g_{rs}\,
 {\dot \eta}_1^r(\tau)\, {\dot \eta}^s_1(\tau)]}(\tau, \sigma^u)
 +\nonumber \\
 &+&{{\delta^3(\sigma^u - \eta^u(\tau))}\over {\mu(\tau)}}\,
 {{g_{\tau\tau} + 2\, g_{\tau r}\, {\dot \eta}^r(\tau) + g_{rs}\,
 {\dot \eta}^r(\tau)\, {\dot \eta}^s(\tau)}\over {\sgn\, l_{\mu}\,
 z^{\mu}_{\tau}}}(\tau, \sigma^u) \Big).
 \label{2.8}
 \eea
\medskip

Besides Eqs.(\ref{2.2}) there is the new momentum

\beq
 \kappa_{1r}(\tau)\, =\, - \sgn\, m_1\, c\, {{g_{\tau r}(\tau, \eta^u_1(\tau)) +
 g_{rs}(\tau, \eta_1^u(\tau))\, {\dot \eta}_1^s(\tau)}\over
 {\sqrt{\sgn\, [g_{\tau\tau}(\tau, \eta^u_1(\tau)) + 2\,
 g_{\tau r}(\tau, \eta^u_1(\tau))\, {\dot \eta}_1^r(\tau) +
 g_{rs}(\tau, \eta^u_1(\tau))\,
 {\dot \eta}_1^r(\tau)\, {\dot \eta}^s_1(\tau)]}}},
 \label{2.9}
 \eeq

\noindent and the first class constraints (\ref{2.3}) become

\bea
 {\cal H}_{\mu}(\tau, \sigma^u) &=& \rho_{\mu}(\tau, \sigma^u) -
 \nonumber \\
 &-& l_{\mu}(\tau, \sigma^u)\, \Big[\delta^3(\sigma^u - \eta^u(\tau))\,
 \Big({1\over {\mu(\tau)}} + {{\mu(\tau)}\over 4}\, h^{rs}(\tau,
 \sigma^u)\, \kappa_r(\tau)\, \kappa_s(\tau)\Big) +\nonumber \\
 &+& \delta^3(\sigma^u -
 \eta_1^u(\tau))\, \sqrt{m_1^2\, c^2 + h^{rs}(\tau, \sigma^u)\,
 \kappa_{1r}(\tau)\, \kappa_{1s}(\tau)}\Big] +\nonumber \\
 &+&\sgn\, \Big(z^{\mu}_r\, h^{rs}\Big)(\tau, \sigma^u)\, \Big[\delta^3(\sigma^u -
 \eta^u(\tau))\, \kappa_s(\tau) + \delta^3(\sigma^u - \eta_1^u(\tau))\,
 \kappa_{1s}(\tau)\Big] \approx 0.
 \label{2.10}
 \eea
\medskip

Eqs.(\ref{2.4}) and (\ref{2.5}) are not changed, so that the final
form of Eq.(\ref{2.7}) is

\bea
  {\cal H}_{\mu}(\tau, \sigma^u) &=& \rho_{\mu}(\tau, \sigma^u) -
  l_{\mu}(\tau, \sigma^u)\, \Big[\delta^3(\sigma^u - \eta^u(\tau))\,
 \sqrt{h^{rs}(\tau, \sigma^u)\, \kappa_r(\tau)\, \kappa_s(\tau)}
 +\nonumber \\
 &+& \delta^3(\sigma^u - \eta_1^u(\tau))\, \sqrt{m_1^2\, c^2 + h^{rs}(\tau,
 \sigma^u)\, \kappa_{1r}(\tau)\, \kappa_{1s}(\tau)}\Big] +\nonumber \\
 &+&\sgn\, \Big(z^{\mu}_r\, h^{rs}\Big)(\tau, \sigma^u)\, \Big[\delta^3(\sigma^u -
 \eta^u(\tau))\, \kappa_s(\tau) + \delta^3(\sigma^u - \eta_1^u(\tau))\,
 \kappa_{1s}(\tau)\Big] \approx 0.\nonumber \\
 &&{}
 \label{2.11}
 \eea

\bigskip

Therefore we get

\bea
 P^{\mu} &=& \int d^3\sigma\, \rho^{\mu}(\tau, \sigma^u) \approx
 \nonumber \\
 &\approx& l^{\mu}(\tau, \eta^u(\tau))\, \sqrt{h^{rs}(\tau, \eta^u(\tau))\,
 \kappa_r(\tau)\, \kappa_s(\tau)} - \sgn\, \Big(z^{\mu}_r\, h^{rs}\Big)(\tau,
 \eta^u(\tau))\, \kappa_s(\tau) +\nonumber \\
 &+&  l^{\mu}(\tau, \eta_1^u(\tau))\, \sqrt{m_1^2\, c^2 + h^{rs}(\tau, \eta_1^u(\tau))\,
 \kappa_{1r}(\tau)\, \kappa_{1s}(\tau)} - \sgn\, \Big(z^{\mu}_r\, h^{rs}\Big)(\tau,
 \eta_1^u(\tau))\, \kappa_{1s}(\tau) =\nonumber \\
 &{\buildrel {def}\over =}& P^{\mu}_o + P^{\mu}_1.
 \label{2.12}
 \eea
\medskip

Since we have $\sgn\, P^2_o = 0$ and $\sgn\, P^2_1 = m_1^2\, c^2$ so
that $P^{\mu}_o$ is a future-pointing null 4-vector and $P^{\mu}_1$
a future-pointing time-like one, we have $\sgn\, P_o \cdot P_1 > 0$.
Therefore we get $\sgn\, P^2 = m_1^2\, c^2 + 2\, \sgn\, P_o \cdot
P_1 > 0$. Let us remark that in our approach the 4-vectors
$P^{\mu}_o$ and $P^{\mu}_1$ are not canonical momenta but derived
quantities.
\bigskip

This implies the following rest-frame instant form description of
the two particles on the Wigner instantaneous 3-space
$z^{\mu}_W(\tau, \vec \sigma) = Y^{\mu}(\tau) +
\epsilon^{\mu}_r(\vec h)\, \sigma^r$

\begin{eqnarray*}
 x^{\mu}(\tau) &=& z^{\mu}_W(\tau, \vec \eta(\tau)) = Y^{\mu}(0) +
 h^{\mu}\, \tau + \epsilon^{\mu}_r(\vec h)\, \eta^r(\tau),\nonumber \\
 &&{}\nonumber \\
 &&{\dot x}^{\mu}(\tau) = h^{\mu} + \epsilon^{\mu}_r(\vec h)\, {\dot
 \eta}^r(\tau),\quad {\dot x}^2(\tau) = 0,\quad {\dot {\vec
 \eta}}^2(\tau) = 1,\nonumber \\
 &&{}\nonumber \\
 P^{\mu}_o &=& h^{\mu}\, \sqrt{{\vec \kappa}^2(\tau)} - \epsilon^{\mu}_r(\vec h)\,
 \kappa_r(\tau),\quad P^2_o = 0,
\end{eqnarray*}

\bea
  x^{\mu}_1(\tau) &=& z^{\mu}_W(\tau, {\vec \eta}_1(\tau)) = Y^{\mu}(0) +
 h^{\mu}\, \tau + \epsilon^{\mu}_r(\vec h)\, \eta^r_1(\tau),\nonumber \\
 &&{}\nonumber \\
 &&{\dot x}_1^{\mu}(\tau) = h^{\mu} + \epsilon^{\mu}_r(\vec h)\, {\dot
 \eta}^r_1(\tau),\quad \sgn\, {\dot x}^2_1(\tau) = 1 - {\dot {\vec
 \eta}}_1^2(\tau) > 0,\nonumber \\
 &&{}\nonumber \\
 P^{\mu}_1 &=& h^{\mu}\, \sqrt{m_1^2\, c^2 + {\vec \kappa}^2_1(\tau)} -
 \epsilon^{\mu}_r(\vec h)\, \kappa_{1r}(\tau),\quad \sgn\, P^2_1 = m_1^2\, c^2,
 \label{2.13}
  \eea

\bigskip

From Ref.\cite{5}, as said in the Introduction, we get the following
form of the external Poincare' generators

\bea
 P^{\mu} &=& M\, c\, h^{\mu} = M\, c\, \Big(\sqrt{1 + {\vec h}^2};
 \vec h\Big),\nonumber \\
 &&{}\nonumber \\
 J^{ij} &=& z^i\, h^j - z^j\, h^i + \epsilon^{ijk}\, S^k,\qquad
 K^i = J^{oi} = - \sqrt{1 + {\vec h}^2}\, z^i + {{(\vec S \times
 \vec h)^i}\over {1 + \sqrt{1 + {\vec h}^2}}},
 \label{2.14}
 \eea

\noindent and the following form of the internal Poincare'
generators

\bea
 M\, c &=& {1\over c}\, {\cal E}_{(int)} = \sqrt{{\vec \kappa}^2} + \sqrt{m_1^2\, c^2 + {\vec
 \kappa}^2_1},\nonumber \\
 {\vec {\cal P}}_{(int)} &=& \vec \kappa + {\vec \kappa}_1 \approx
 0,\nonumber \\
 \vec S &=& {\vec {\cal J}}_{(int)} = \vec \eta \times \vec \kappa +
 {\vec \eta}_1 \times {\vec \kappa}_1,\nonumber \\
 {\vec {\cal K}}_{(int)} &=& - \vec \eta\, \sqrt{{\vec \kappa}^2} -
 {\vec \eta}_1\, \sqrt{m_1^2\, c^2 + {\vec \kappa}_1^2} \approx 0.
 \label{2.15}
 \eea

\medskip

As shown in Refs.\cite{5,7,8,9}, the explicit $\tau$-dependence of
the gauge fixing $z^{\mu}(\tau, \sigma^r) - z_W^{\mu}(\tau,
\sigma^r) \approx 0$, defining the inertial rest-frame instant form,
implies that the Dirac Hamiltonian $H_D \approx 0$ of Eq.(\ref{2.4})
is replaced by the internal invariant mass $Mc = {1\over c}\, {\cal
E}_{(int)} = \sqrt{\sgn\, P^2}$ as the effective Hamiltonian for the
evolution of the internal variables inside the Wigner 3-space.

\vfill\eject

\section{The Pseudo-Classical Photon: a Massless Particle with
Helicity}

To describe the helicity of a pseudo-classical photon we follow the
method of Ref.\cite{15}. Let us associate two complex Grassmann
4-vector $\theta^{* \mu}(\tau)$, $\theta^{\mu}(\tau)$ ($\theta^{*
\mu}\, \theta^{* \nu} + \theta^{* \nu}\, \theta^{* \mu} = \theta^{
\mu}\, \theta^{ \nu} + \theta^{ \nu}\, \theta^{ \mu} = \theta^{
\mu}\, \theta^{* \nu} + \theta^{* \nu}\, \theta^{ \mu} = 0$,
$(\theta^{* \mu})^2 = (\theta^{\mu})^2 = 0$ for each value of $\mu$)
to the massless particle.\medskip

If $\zeta^*(\tau)$, $\zeta(\tau)$ are Grassmann Lagrange multipliers
and $\nu(\tau)$ a Lagrange multiplier, the action (\ref{2.8}) is
replaced by the following one

\bea
  S &=& \int d\tau\, d^3\sigma\, \Big(- m_1\, c\, \delta^3(\sigma^u - \eta^u_1(\tau))\,
 \sqrt{\sgn\, [g_{\tau\tau} + 2\, g_{\tau r}\, {\dot \eta}_1^r(\tau) + g_{rs}\,
 {\dot \eta}_1^r(\tau)\, {\dot \eta}^s_1(\tau)]}(\tau, \sigma^u)
 +\nonumber \\
 &+&{{\delta^3(\sigma^u - \eta_1^u(\tau))}\over {\mu(\tau)\, \sgn\,
 l_{\mu}\, \Big[z^{\mu}_{\tau}(\tau, \sigma^u) + \zeta^*\, \theta^{* \mu} -
 \zeta\, \theta^{\mu}\Big]}}\nonumber \\
 &&\eta_{\mu\nu}\, \Big[z^{\mu}_{\tau} + z^{\mu}_r\, {\dot \eta}^r(\tau)
 + \zeta^*(\tau)\, \theta^{* \mu}(\tau) - \zeta(\tau)\, \theta^{\mu}(\tau)\Big](\tau,
 \sigma^u)\nonumber \\
 &&\Big[z^{\nu}_{\tau} + z^{\nu}_s\, {\dot \eta}^s(\tau)
 + \zeta^*(\tau)\, \theta^{* \nu}(\tau) -
 \zeta(\tau)\, \theta^{\nu}(\tau)\Big](\tau, \sigma^u)
 -\nonumber \\
 &-&\delta^3(\sigma^u - \eta_1^u(\tau))\, \Big[{i\over 2}\,
 \Big(\theta^{* \mu}(\tau)\, {\dot \theta}_{\mu} -
 {\dot \theta}^{* \mu}(\tau)\, \theta_{\mu}(\tau)\Big) + \nu(\tau)\, \theta^{*
 \mu}(\tau)\, \theta_{\mu}(\tau)\Big].
 \label{3.1}
 \eea

\bigskip

The Grassmann momenta are $\Pi^{\mu}_{\theta} = - {{\partial\,
L}\over {\partial\, {\dot \theta}_{\mu}}} = - {i\over 2}\, \theta^{*
\mu}$, $\Pi^{\mu}_{\theta^*} = - {{\partial\, L}\over {\partial\,
{\dot \theta}^*_{\mu}}} = - {i\over 2}\, \theta^{\mu}$ with Poisson
brackets $\{ \Pi_{\theta \mu}, \theta^{\nu}\} = \{ \Pi_{\theta^*
\mu}, \theta^{* \nu}\} = \eta^{\nu}_{\mu}$. Since they imply second
class constraints, the Grassmann momenta can be eliminated by using
the Dirac brackets $\{ \theta^{\mu}, \theta^{* \nu}\} = i\,
\eta^{\mu\nu}$ (we use the notation $\{ .,. \}$ also for $\{ .,.
\}^*$ for simplicity).

\bigskip

Besides Eq.(\ref{2.9}) the other canonical momenta are

\begin{eqnarray*}
  \kappa_r(\tau) &=& {{\partial\, L(\tau)}\over {\partial\, {\dot \eta}^r}}
 =  - {{2}\over {\mu(\tau)}}\, {{N_r + h_{rs}\, {\dot \eta}^s(\tau)
  - \sgn\, z_{r\mu}\, (\zeta^*\, \theta^{* \mu} - \zeta\, \theta^{\mu})}\over
  {l_{\mu}\, \Big[z^{\mu}_{\tau} + \zeta^*\, \theta^{* \mu} -
 \zeta\, \theta^{\mu}\Big]}} (\tau, \eta^u(\tau)),\nonumber \\
 &&{}\nonumber \\
 \rho_{\mu}(\tau, \sigma^u) &=& {{\delta\, S}\over {\delta\,
 z^{\mu}_{\tau}(\tau, \sigma^u)}} =
  {{\delta^3(\sigma^u - \eta^u (\tau))}
 \over {\mu(\tau)\, \sgn\,  \Big(l_{\mu}\, \Big[z^{\mu}_{\tau} + \zeta^*\,
 \theta^{* \mu} - \zeta\, \theta^{\mu}\Big]\Big)(\tau, \sigma^v)}}
 \nonumber \\
 &&\Big[2\, (z_{\tau \mu} + z_{r \mu}\, {\dot \eta}^r(\tau) + \zeta^*\,
 \theta^{*}_{\mu} - \zeta\, \theta_{\mu}) -\nonumber \\
  &-& {{ l_{\mu}}\over {l_{\mu}\, \Big[z^{\mu}_{\tau} + \zeta^*\,
 \theta^{* \mu} - \zeta\, \theta^{\mu}\Big]}}\, \eta_{\alpha\beta}
 \nonumber \\
 &&\Big(z_{\tau}^{\alpha} + z_{r}^{\alpha}\, {\dot \eta}^r(\tau) + \zeta^*\,
 \theta^{* \alpha} - \zeta\, \theta^{\alpha}\Big)\nonumber \\
 &&\Big(z_{\tau}^{\beta} + z_{r}^{\beta}\, {\dot \eta}^r(\tau) +
 \zeta^*\, \theta^{* \beta} - \zeta\, \theta^{\beta}\Big)
 \Big](\tau, \sigma^v),
 \end{eqnarray*}

 \bea
 \pi(\tau) &=& {{\partial\, L(\tau)}\over {\partial\, {\dot \mu}}} = 0,
 \nonumber \\
 \pi_{\zeta}(\tau) &=& {{\partial\, L(\tau)}\over {\partial\,
 {\dot \zeta}}} = 0,\qquad
 \pi_{\zeta^*}(\tau) = {{\partial\, L(\tau)}\over {\partial\,
 {\dot \zeta^*}}} = 0,\nonumber \\
 \pi_{(\nu )}(\tau) &=& {{\partial\, L(\tau)}\over {\partial\,
 {\dot \nu}}} = 0.
 \label{3.2}
 \eea

\bigskip

The primary constraints are $\pi(\tau) \approx 0$,
$\pi_{\zeta}(\tau) \approx 0$, $\pi_{\zeta^*}(\tau) \approx 0$,
$\pi_{(\nu)}(\tau) \approx 0$, and the contraints ${\cal H}_\mu(\tau,\vec{\sigma})\approx 0$,
given in eq.(\ref{2.10}).

\medskip

The Dirac Hamiltonian is
\bea
 H_D &=& \int d^3\sigma\, \lambda^{\mu}(\tau, \sigma^u)\, {\cal
 H}_{\mu}(\tau, \sigma^u) +\nonumber \\
 &+& \lambda(\tau)\, \pi(\tau) + \lambda_{(\nu)}(\tau)\, \pi_{(\nu)}(\tau)
 + \lambda_{\zeta}(\tau)\, \pi_{\zeta}(\tau) + \lambda_{\zeta^*}(\tau)\,
 \pi_{\zeta^*}(\tau) +\nonumber \\
 &+& \zeta^*(\tau)\, \psi^*(\tau) + \zeta(\tau)\, \psi(\tau) +
 \nu(\tau)\, \theta^*_{\mu}(\tau)\, \theta^{\mu}(\tau),
 \label{3.4}
 \eea
where
\bea
 &&\theta^*_{\mu}(\tau)\, \theta^{\mu}(\tau) \approx\, 0,\nonumber \\
 &&{}\nonumber \\
 &&\psi(\tau) = \Big[l_{\mu}\, \Big({1\over {\mu(\tau)}} + {{\mu(\tau)}\over 4}\,
 h^{rs}\, \kappa_r(\tau)\, \kappa_s(\tau)\Big) - \sgn\, z_{r \mu}\,
 h^{rs}\, \kappa_s(\tau)\Big](\tau, \eta^u(\tau))\, \theta^{\mu}(\tau)
 \approx\, 0,\nonumber \\
 &&\psi^*(\tau) = \Big[l_{\mu}\, \Big({1\over {\mu(\tau)}} + {{\mu(\tau)}\over 4}\,
 h^{rs}\, \kappa_r(\tau)\, \kappa_s(\tau)\Big) - \sgn\, z_{r \mu}\,
 h^{rs}\, \kappa_s(\tau)\Big](\tau, \eta^u(\tau))\, \theta^{* \mu}(\tau)
 \approx\, 0.\nonumber \\
 &&
 \label{3.3}
 \eea
are secondary constraints implied by the $\tau$-preservation of the primary constraints
$\pi_{\zeta}(\tau) \approx 0$, $\pi_{\zeta^*}(\tau) \approx 0$,
$\pi_{(\nu)}(\tau) \approx 0$.

\medskip

\noindent In obtaining this result we used the following consequence
of the Grassmann nature of the variables: $\Big[l_{\mu}\,
\Big(\zeta^*\, \theta^{* \mu} - \zeta\, \theta^{\mu}\Big)\Big]^2 =
0$.

\medskip

The $\tau$-preservation of the primary constraints implies the
secondary constraint of Eq.(\ref{2.5}), so that we can eliminate the
pair of variables $\mu(\tau)$ and $\pi(\tau)$.

\bigskip

The final form of the first class constraints is

\bea
   {\cal H}_{\mu}(\tau, \sigma^u) &=& \rho_{\mu}(\tau, \sigma^u) -
  l_{\mu}(\tau, \sigma^u)\, \Big[\delta^3(\sigma^u - \eta^u(\tau))\,
 \sqrt{h^{rs}(\tau, \sigma^u)\, \kappa_r(\tau)\, \kappa_s(\tau)}
 +\nonumber \\
 &+& \delta^3(\sigma^u - \eta_1^u(\tau))\, \sqrt{m_1^2\, c^2 + h^{rs}(\tau,
 \sigma^u)\, \kappa_{1r}(\tau)\, \kappa_{1s}(\tau)}\Big] +\nonumber \\
 &+&\sgn\, \Big(z^{\mu}_r\, h^{rs}\Big)(\tau, \sigma^u)\, \Big[\delta^3(\sigma^u -
 \eta^u(\tau))\, \kappa_s(\tau) + \delta^3(\sigma^u - \eta_1^u(\tau))\,
 \kappa_{1s}(\tau)\Big] \approx 0,\nonumber \\
 &&{}\nonumber \\
  &&\theta^*_{\mu}(\tau)\, \theta^{\mu}(\tau) \approx\, 0,\nonumber \\
 &&{}\nonumber \\
 &&\psi(\tau) = \Big[l_{\mu}\, \sqrt{h^{rs}\, \kappa_r(\tau)\, \kappa_s(\tau)}
  - \sgn\, z_{r \mu}\, h^{rs}\, \kappa_s(\tau)\Big](\tau, \eta^u(\tau))\,
  \theta^{\mu}(\tau) \approx\, 0,\nonumber \\
 &&\psi^*(\tau) = \Big[l_{\mu}\, \sqrt{h^{rs}\, \kappa_r(\tau)\, \kappa_s(\tau)} -
 \sgn\, z_{r \mu}\, h^{rs}\, \kappa_s(\tau)\Big](\tau, \eta^u(\tau))\,
 \theta^{* \mu}(\tau) \approx\, 0,\nonumber \\
 &&{}\nonumber \\
 \pi_{\zeta}(\tau) &\approx& 0,\qquad
  \pi_{\zeta^*}(\tau) \approx 0,\qquad
  \pi_{(\nu)}(\tau) \approx 0.
 \label{3.5}
 \eea

The Dirac Hamiltonian (\ref{3.4}) becomes $H_D = \int d^3\sigma\,
\lambda^{\mu}(\tau, \sigma^u)\, {\cal H}_{\mu}(\tau, \sigma^u) +
\lambda_{(\nu)}(\tau)\, \pi_{(\nu)}(\tau) + \lambda_{\zeta}(\tau)\,
\pi_{\zeta}(\tau) + \lambda_{\zeta^*}(\tau)\, \pi_{\zeta^*}(\tau) +
\zeta^*(\tau)\, \psi^*(\tau) + \zeta(\tau)\, \psi(\tau) +
 \nu(\tau)\, \theta^*_{\mu}(\tau)\, \theta^{\mu}(\tau)$. If we add
the gauge fixings $\zeta(\tau) \approx 0$, $\zeta^*(\tau) \approx
0$, $\nu(\tau) \approx 0$, whose $\tau$-preservation implies
$\lambda_{\zeta^*}(\tau) = \lambda_{\zeta}(\tau) =
\lambda_{(\nu)}(\tau) = 0$, we can eliminate these variables and
their conjugate momenta. Then the final Dirac Hamiltonian is

\beq
 H_D = \int d^3\sigma\, \lambda^{\mu}(\tau, \sigma^u)\, {\cal
H}_{\mu}(\tau, \sigma^u).
 \label{3.6}
 \eeq
\medskip

Let us remark that often in the literature one use an extended
Hamiltonian

 \bea
 H_E &=& \int d^3\sigma\, \lambda^{\mu}(\tau, \sigma^u)\, {\cal
 H}_{\mu}(\tau, \sigma^u) + \gamma^*(\tau)\, \psi^*(\tau) + \gamma(\tau)\, \psi(\tau) +
 \beta (\tau)\, \theta^*_{\mu}(\tau)\, \theta^{\mu}(\tau),
 \label{3.a}
 \eea

\noindent which takes into account all the constraints of
Eqs.(\ref{3.3}).

\bigskip

In the rest-frame instant form on the Wigner instantaneous 3-spaces
with $z^{\mu}(\tau, \sigma^r) = z_W^{\mu}(\tau, \sigma^r)$, where
$l^{\mu}(\tau, \sigma^u) = h^{\mu}$, the constraints on the $\theta$
variables can be rewritten in the form

\beq
 \psi^*(\tau) = P_{o \mu}\, \theta^{* \mu}(\tau) \approx 0,\qquad
 \psi(\tau) = P_{o\, \mu}\,
 \theta^{\mu}(\tau) \approx 0,\qquad \theta^*_{\mu}(\tau)\, \theta^{\mu}(\tau)
 \approx 0,
 \label{3.7}
 \eeq

\noindent where $P^{\mu}_o$ was defined in Eq.(\ref{2.13}).\medskip

Since the Grassmann variables are Lorentz 4-vectors, the ten
Poincare' generators generated by the action (\ref{3.1}) are now
$P^{\mu} = \int d^3\sigma\, \rho^{\mu}(\tau, \sigma^u)$ and
$J^{\mu\nu} = \int d^3\sigma\, \Big(z^{\mu}\, \rho^{\nu} - z^{\nu}\,
\rho^{\mu}\Big)(\tau, \sigma^u) + S^{\mu\nu}_{\theta}$ with

\beq
 S^{\mu\nu}_{\theta} = - i\, (\theta^{*\, \mu}\, \theta^{\nu} -
\theta^{*\, \nu}\, \theta^{\mu}),\qquad P_{o\, \mu}\,
S^{\mu\nu}_{\theta} \approx 0.
 \label{3.8}
 \eeq

\bigskip

As shown in Ref.\cite{15},  the gauge fixings to the transversality
constraints $\psi^*(\tau) \approx 0$ and $\psi(\tau) \approx 0$ are

\beq
 \phi^*(\tau) = K_o(P_o) \cdot \theta^* \approx 0,\qquad
 \phi(\tau) = K_o(P_o) \cdot \theta \approx 0,
 \label{3.9}
 \eeq

\noindent where the null 4-vector $K^{\mu}_o(P_o)$ is defined in
Appendix A.
\medskip

In this way we get two pairs of second class constraints with the
following Poisson brackets

\bea
 &&\{ \psi, \phi^* \} = \{ \psi^*, \phi \} = i\, P_o \cdot K_o(P_o)
 = i,\qquad \{ \psi, \phi \} = \{ \psi^*, \phi^* \} = 0,\nonumber \\
 &&\{ \theta^* \cdot \theta, \phi \} = - i\, \phi \approx 0,\qquad
 \{ \theta^* \cdot \theta, \phi^* \} = - i\, \phi^* \approx 0,
 \label{3.10}
 \eea

\noindent so that the elimination of these constraints implies the
following Dirac brackets \cite{15}

\bea
 &&\{ A, B \}^* = \{ A, B \} + {i\over {P_o \cdot K_o(P_o)}}\,
 \Big[\{ A, \psi \}\, \{ \phi^*, B \} +\nonumber \\
 &+& \{ A, \psi^*\}\, \{\phi, B \} + \{ A, \phi \}\, \{ \psi^* \} +
 \{ A, \phi^*\}\, \{ \psi, B \} -\nonumber \\
 &-& {{P^2_o}\over {P_o \cdot K_o(P_o)}}\, \Big(\{ A, \phi \}\, \{
 \phi^*, B \} + \{ A, \phi^*\}\, \{ \phi, B \} \Big)\Big].
 \label{3.11}
 \eea

\medskip

As shown in Ref.\cite{15}, the variables $\theta^{* \mu}(\tau)$ and
$\theta^{\mu}(\tau)$ can be now replaced by the following ones

\bea
 &&{\tilde \theta}_{\lambda}(\tau) = \theta_{\mu}(\tau)\,
 \epsilon^{\mu}_\lambda(P_o),\qquad {\tilde \theta}^*_{\lambda}(\tau) =
 \theta^*_{\mu}(\tau)\, \epsilon^{\mu}_\lambda(P_o),\quad \lambda =1,2,
 \nonumber \\
 &&{}\nonumber \\
 &&\qquad \{ {\tilde \theta}_{\lambda}, {\tilde \theta}^*_{\lambda^{'}} { \} }^* = - i\,
 \delta_{\lambda\lambda^{'}},\nonumber \\
 &&{}\nonumber \\
 && \theta^*(\tau) \cdot \theta(\tau) = - \sum_{\lambda =1}^2\,
 {\tilde \theta}^*_{\lambda}(\tau)\, {\tilde
 \theta}_{\lambda}(\tau) \approx 0,
 \label{3.12}
 \eea

\noindent where the polarization vectors
$\epsilon^{\mu}_{\lambda}(P_o)$ are defined in  Eqs.(\ref{a3}).

\medskip

However, since $P_o^{\mu}$ depends on the momenta $\vec
\kappa(\tau)$ and $\vec h$, also the variables $\vec \eta(\tau)$ of
the massless particle and the Jacobi data $\vec z$ have to be
modified

\bea
 {\vec z}^{'} &=&\vec{z}+
\frac{1}{2}\left[\,P_{o\sigma}\frac{\partial
P_{o\rho}}{\partial\vec{h}}+ K_{o\sigma}\frac{\partial
K_{o\rho}}{\partial\vec{h}}-\sum_\lambda\,
\epsilon_{\lambda\sigma}\frac{\partial
\epsilon_{\lambda\rho}}{\partial\vec{h}}\right]\,
S_{\theta}^{\rho\sigma},\qquad \{ z^{{'}\, i}, h^j \}^* =
 \delta^{ij},\nonumber \\
 &&{}\nonumber \\
 {\vec \eta}^{'} &=&\vec{\eta}+
\frac{1}{2}\left[\,P_{o\sigma}\frac{\partial P_{o\rho}}{\partial\vec{\kappa}}+
K_{o\sigma}\frac{\partial K_{o\rho}}{\partial\vec{\kappa}}-\sum_\lambda\,
\epsilon_{\lambda\sigma}\frac{\partial \epsilon_{\lambda\rho}}{\partial\vec{\kappa}}\right]\,
S_{\theta}^{\rho\sigma},\qquad \{ \eta^{{'}\, r}, \kappa^s
 \}^* = \delta^{rs}.
 \label{3.13}
 \eea

 \bigskip

Eqs.(\ref{3.12}) imply ($\epsilon_{\lambda\lambda^{'}} = -
\epsilon_{\lambda^{'}\lambda}$, $\epsilon_{12} = 1$)

\bea
 S^{\mu\nu}_{\theta} &=& \sum_{\lambda\lambda^{'}}\, \epsilon^{\mu}_{\lambda}(P_o)
 \, \epsilon^{\nu}_{\lambda^{'}}(P_o)\, S_{\lambda\lambda^{'}},\nonumber \\
 &&{}\nonumber \\
 &&S_{\lambda\lambda^{'}} = - i\, ({\tilde \theta}^*_{\lambda}\, {\tilde
 \theta}_{\lambda^{'}} - {\tilde \theta}^*_{\lambda^{'}}\, {\tilde
 \theta}_{\lambda}) = \epsilon_{\lambda\lambda^{'}}\,
 \Sigma,\nonumber \\
 &&\Sigma = S_{\theta \mu\nu}\, \epsilon^{\mu}_1(P_o)\,
 \epsilon^{\nu}_2(P_o) = - i\, ({\tilde \theta}^*_1\, {\tilde
 \theta}_2 - {\tilde \theta}^*_2\, {\tilde \theta}_1).
 \label{3.14}
 \eea

 \bigskip

Now we have $S^{\mu o}_{\theta} = 0$ and ($S^i_{\theta} = {1\over
2}\, \epsilon^{ijk}\, S_{\theta}^{jk}$)

\beq
 {\vec S}_{\theta} = {{{\vec P}_o}\over {|{\vec P}_o|}}\, \Sigma,
 \qquad \Sigma = {{{\vec P}_o \cdot {\vec S}_{\theta}}\over {|{\vec
 P}_o|}}.
 \label{3.15}
 \eeq

Therefore $\Sigma$ describes the helicity of the massless photon,
which has the spin collinear with the 3-momentum.

 \bigskip

The $\tau$-preservation of the gauge fixings (\ref{3.9}) implies the
vanishing of the corresponding Dirac multipliers $\gamma^*(\tau) =
\gamma(\tau) = 0$ in the Hamiltonian $H_E$ of Eq.(\ref{3.a}), which
in the inertial rest frame is replaced by the effective Hamiltonian

\bea
 H &=& M c - \beta(\tau)\, \sum_{\lambda}\, \theta^*_{\lambda}(\tau)\,
 \theta_{\lambda}(\tau),\nonumber \\
 &&{}\nonumber \\
 M c &=& {1\over c}\, {\cal E}_{(int)} = \sqrt{{\vec \kappa}^2} +
 \sqrt{m_1^2\, c^2 + {\vec \kappa}_1^2}.
 \label{3.16}
 \eea

\medskip

The external Poincare' generators have the form of Eq.(\ref{2.14})
with $M c$ of Eq.(\ref{3.16}) and internal spin

\beq
 \vec S = {\vec {\cal J}}_{(int)} = \vec \eta \times \vec \kappa +
 {\vec \eta}_1 \times {\vec \kappa}_1 + {\vec S}_{\theta}.
 \label{3.17}
 \eeq

 \medskip

The other internal Poincare' generators, explicitly satisfying the
Poincare' algebra, are

 \bea
 {\vec {\cal P}}_{(int)} &=& \vec \kappa + {\vec \kappa}_1 \approx
 0,\nonumber \\
 {\vec {\cal K}}_{(int)} &=& - \vec \eta\, \sqrt{{\vec \kappa}^2} -
 {\vec \eta}_1\, \sqrt{m_1^2\, c^2 + {\vec \kappa}_1^2} + {\vec
 S}_{\theta} \approx 0.
 \label{3.18}
 \eea

\noindent Their vanishing eliminates the internal 3-center of mass
inside the Wigner 3-space and its conjugate momentum. The form of
the internal Lorentz boosts, with the correct Poisson brackets with
the other generators, has been guessed, since the action (\ref{3.1})
contains Lagrange multipliers which make difficult to find an
energy-momentum tensor independent from them.

\vfill\eject

\section{Quantization}

Let us add some remarks about the quantization of the rest-frame
instant form.

The quantization of the bosonic variables $\vec z$, $\vec h$, ${\vec
\eta}_1$, ${\vec \kappa}_1$, and $\vec \eta$, $\vec \kappa$ is done
with the new rest-frame quantization scheme for relativistic quantum
mechanics introduced in Ref.\cite{16} \footnote{See Ref.\cite{17}
for the quantization of spinning massive particles} in an
un-physical Hilbert space ${\cal H} = {\cal H}_{com} \otimes {\cal
H}_{{\vec \eta}_1} \otimes {\cal H}_{\vec \eta}$, where ${\cal
H}_{com}$ is the Hilbert space of the frozen Jacobi data of the
external decoupled center of mass. The physical states and the
associated physical Hilbert space have to be identified by solving
the restrictions $< \Phi_{phys} | {\hat {\vec {\cal P}}}_{(int)}\, |
\Phi_{phys} > \, =\, < \Phi_{phys} | {\hat {\vec {\cal
K}}}_{(int)}\, | \Phi_{phys} > \, = 0$ (quantization of the
rest-frame conditions (\ref{3.18}) eliminating the internal center
of mass inside the Wigner 3-space). The resulting physical Hilbert
space should have the structure ${\cal H}_{(phys)} = {\cal H}_{com}
\otimes {\cal H}_{rel}$, where ${\cal H}_{rel}$ is the internal
Hilbert space associated to the Wigner-covariant relative variables
${\vec \rho}_{12}$, ${\vec \pi}_{12}$, describing the isolated
system of the massive plus massless particles in the rest frame.
\medskip

As shown in Ref.\cite{16} the quantization of the frozen Jacobi data
$\vec z$, $\vec h$, is done in the preferred $\vec h$-basis
(definition of the rest frame) in the momentum representation: $h^i
\mapsto h^i$, $z^i \mapsto i \hbar\, {{\partial}\over {\partial\,
h^i}} - i \hbar\, {{h^i}\over {1 + {\vec h}^2}}$. The center-of-mass
wave function with 3-velocity $\vec k$ is $\psi_{\vec k}(\vec h) =
\delta^3(\vec h - \vec k)$ and the scalar product is $< \psi_1 |
\psi_2 > = \int {{d^3h}\over {2\, \sqrt{1 + {\vec h}^2}}}\,
\psi_1^*(\vec h)\, \psi_2(\vec h)$.\medskip

The unphysical Hilbert space ${\cal H}_{{\vec \eta}_1}$, with scalar
product $<\phi_1, \phi_2> = \int d^3\eta_1\, \phi_1^*(\tau, {\vec
\eta}_1)\, \phi_2(\tau, {\vec \eta}_1)$ in the coordinate
representation, is the standard space of a massive particle whose
positive energy operator $\sqrt{m_1^2\, c^2 + {\hat {\vec
\kappa}}_1}$ is a pseudo-differential operator defined in
Ref.\cite{q1}.\medskip

Instead the unphysical Hilbert space ${\cal H}_{\vec \eta}$ of the
massless particle is a suitable limit for zero mass of the precedent
Hilbert space. The delicate point is to see whether the methods of
Ref.\cite{q2} (with a smoothing around $\vec \kappa = 0$) allow one
to define a massless pseudo-differential operator $\sqrt{{\hat {\vec
\kappa}}^2}$, such that the velocity operator ${d\over {d\tau}}\,
{\hat \eta}^r(\tau ) = [{\hat \eta}^r(\tau), \sqrt{{\hat {\vec
\kappa}}^2}]$ is well defined and satisfies $\sum_r\, ({d\over
{d\tau}}\, {\hat \eta}^r(\tau ))^2 = 1$.\medskip

If this problem has a well-defined solution and if the physical
Hilbert space ${\cal H}_{rel}$ with its scalar product can be
identified, then the physical Hamiltonian and the rest spin will be
operators depending only on the relative variables ${\vec
\rho}_{12}$, ${\vec \pi}_{12}$, and corresponding to
Eqs.(\ref{3.16}) and (\ref{3.17}) restricted by Eqs.(\ref{3.18}).
This problem will be studied elsewhere.

\bigskip

Instead we  add the quantization of the Grassmann-valued helicity of
a massless particle by enlarging its Hilbert space ${\cal H}_{\vec
\eta}$ in the following way.
\medskip

As shown in Ref.\cite{15}, the quantization rule $\theta^{\mu}
\mapsto b^{\mu}$, $\theta^{* \mu} \mapsto b^{\dagger \mu}$ leads to
a four-dimensional Fermi oscillator ($[a,b]_+ = ab + ba$ denotes the
anti-commutator)

\beq
 [b^{\mu}, b^{\dagger \nu}]_+ = - \eta^{\mu\nu},\qquad
 [b^{\mu}, b^{\nu}]_+ = [b^{\dagger \mu}, b^{\dagger \nu}]_+ = 0.
 \label{4.1}
 \eeq

Therefore we have a 16-dimensional Hilbert space ${\cal H}_{pol}$,
describing the polarization of the massless particle, spanned by the
basis $|0>$, $|\mu> = b^{\dagger \mu}\, |0>$, $|\mu\nu> = b^{\dagger
\mu}\, b^{\dagger \nu}\, |0>$.

\bigskip

However we have the quantum analogue of the constraints (\ref{3.6})
to take into account: they will restrict ${\cal H}_{pol}$ to a
physical 2-dimensional Hilbert space ${\cal H}_{helicity}$
describing the (1, 0) + (0, 1) helicity representation of the
Poincare' group. The first stage of the reduction is done by
applying the quantum transversality constraints (replacing the
classical ones (\ref{3.7})) with the Gupta-Bleuler method

\beq
 {\hat P}_o \cdot b\, |\psi>_{phys} = 0,\qquad {}_{phys}<\psi|\,
 {\hat P}_o \cdot b^{\dagger} = 0.
 \label{4.1}
 \eeq

\noindent In this way only four states of the basis survive $|0>$,
$A_{\mu}(P_o)\, b^{\dagger \mu}\, |0>$ (with $P_o^{\mu}\,
A_{\mu}(P_o) = 0$) and $F_{\mu\nu}(P_o)\, b^{\dagger \mu}\,
b^{\dagger \nu}\, |0>$ (with $P_o^{\mu}\, F_{\mu\nu}(P_o) = 0$,
$F_{\mu\nu}(P_o) = - F_{\nu\mu}(P_o)$).\medskip

As shown in Ref.\cite{15} the last constraint in Eqs.(\ref{3.12}) is
$(b^{\dagger} \cdot b + \delta)\, |\psi>_{phys} = 0$, where $\delta$
is a c-number present due to ordering problems. To select the
photon-like state $A_{\mu}(P_o)\, b^{\dagger \mu}\, |0>$ (with
$P_o^{\mu}\, A_{\mu}(P_o) = 0$) with two helicity levels one must
choose an ordering corresponding to $\delta = 1$ \cite{15}.

\vfill\eject

\section{Conclusions}

In this paper we have included in the rest-frame instant form of the
dynamics of isolated systems in Minkowski space-time the description
of positive-energy massless particles both without and with
helicity. To avoid the front form of dynamics we must include at
least an additional positive-energy scalar massive particle (or any
other type of matter). This allows to find a classical background
for the particle description of a ray of light in Minkowski
space-time.\medskip

We have also added some comments on how to apply to this isolated
system the procedure of quantization based on the recently developed
rest-frame form of relativistic quantum mechanics \cite{16}. If a
suitable definition for a pseudo-differential operator corresponding
to $\sqrt{{\vec \kappa}^2}$ exists, we have a description of an
isolated {\it photon}, to be compared with the one-particle
approximations used for a photon in Refs.\cite{18}.

\vfill\eject

\appendix

\section{The light-like Polarization Vectors}

By using the future-pointing null vector $P^{\mu}_o = \sqrt{{\vec
\kappa}^2}\, h^{\mu} - \epsilon^{\mu}_r(\vec h)\, \kappa_r =
\Big(P^o_o = |{\vec P}_o|; {\vec P}_o\Big)$ of Eqs.(\ref{2.12}), we
can rewrite Appendix A of Ref.\cite{15} in our formalism.\medskip

For null vectors we have $P^{\mu}_o = L({P_o, {\buildrel \circ \over
P}_o})^{\mu}{}_{\nu}\, {\buildrel \circ \over P}_o^{\nu}$ with
${\buildrel \circ \over P}_o = \omega\, \Big(1; 0, 0, 1\Big)$, where
$L({P_o, {\buildrel \circ \over P}_o})$ is the standard boost for
null vectors whose expression is

\bea
 L({P_o, {\buildrel \circ \over P}_o}) &=&
 \left( \begin{array}{ccc}{1\over 2}\, \Big({{|{\vec P}_o|}\over {\omega}}
 + {{\omega}\over {|{\vec P}_o|}}\Big) & 0 & {1\over 2}\, \Big({{|{\vec
 P}_o|}\over {\omega}} - {{\omega}\over {|{\vec P}_o|}}\Big)\\
 {1\over 2}\, \Big({{|{\vec P}_o|}\over {\omega}} - {{\omega}\over
 {|{\vec P}_o|}}\Big) & \delta^a_{\lambda} + {{P^a_o\, P_{o\, \lambda}}\over
 {|{\vec P}_o|\, (|{\vec P}_o| + P^3_o)}} & {1\over 2}\, \Big({{|{\vec P}_o|}\over {\omega}}
 + {{\omega}\over {|{\vec P}_o|}}\Big) \\
 {1\over 2}\, \Big({{|{\vec P}_o|}\over {\omega}}
 - {{\omega}\over {|{\vec P}_o|}}\Big) & {{P_{o\, \lambda}}\over
 {|{\vec P}_o|}} & {1\over 2}\, \Big({{|{\vec P}_o|}\over {\omega}}
 + {{\omega}\over {|{\vec P}_o|}}\Big)\, {{P^3_o}\over {|{\vec P}_o|}}
 \end{array} \right).
 \label{a1}
 \eea

The light-like polarization vectors are just the columns of this
matrix

\bea
 \epsilon^{\mu}_A(P_o) &=& L({P_o, {\buildrel \circ \over
 P}_o})^{\mu}{}_{\nu}\,\, \epsilon^{\nu}_A({\buildrel \circ\over
 P}_o),\nonumber \\
 &&\eta^{\mu\nu} = \epsilon^{\mu}_A(P_o)\, \eta^{AB}\,
 \epsilon^{\nu}_B(P_o),
 \label{a2}
 \eea

\noindent where $\epsilon^{\mu}_A({\buildrel \circ\over
 P}_o)$ is the standard basis $(1; 0,0,0)$, $(0;1,0,0)$,
 $(0;0,1,0)$, $(0;0,0,1)$.
\medskip

Instead of this basis it is convenient to use the following one

\bea
 P^{\mu}_o &=& \omega\, \Big(\epsilon^{\mu}_o(P_o) + \epsilon^{\mu}_3(P_o)\Big)
 = \Big( |{\vec P}_o|; {\vec P}_o\Big),\nonumber \\
 K^{\mu}_o(P_o) &=&{1\over {2\, \omega}}\, \Big(\epsilon^{\mu}_o(P_o)
 - \epsilon^{\mu}_3(P_o)\Big) = {1\over {2\, |{\vec P}_o|^2}}\,
 \Big( |{\vec P}_o|; -{\vec P}_o\Big),\nonumber \\
 \epsilon^{\mu}_{\lambda}(P_o) &=& \Big(0;\, \delta^a_{\lambda} +
 {{P_o^a\, P_{o\, \lambda}}\over {|{\vec P}_o|\, (|{\vec P}_o| +
 P^3_o)}},\,\,  {{P_{o\, \lambda}}\over {|{\vec P}_o|}}\Big),
 \quad a, \lambda = 1,2,\nonumber \\
 &&{}\nonumber \\
 &&P^2_o = K^2_o(P_o) = 0,\qquad \sgn\, P_o \cdot K_o(P_o) = 1,
 \nonumber \\
 &&P_o \cdot \epsilon_{\lambda}(P_o) = K_o(P_o) \cdot
 \epsilon_{\lambda}(P_o) = 0,\qquad \sgn\, \epsilon_{\lambda}(P_o)
 \cdot \epsilon_{\lambda^{'}}(P_o) = - \delta_{\lambda \lambda^{'}},\nonumber \\
 &&{}\nonumber \\
 &&\eta^{\mu\nu} = P^{\mu}_o\, K^{\nu}_o(P_o) + P^{\nu}_o\,
 K^{\mu}_o(P_o) - \sum_{\lambda}\, \epsilon^{\mu}_{\lambda}(P_o)\,
 \epsilon^{\nu}_{\lambda}(P_o).
 \label{a3}
 \eea

\bigskip

Since we have $P^{\mu}_o = \sqrt{{\vec \kappa}^2}\, h^{\mu} -
\epsilon^{\mu}_r(\vec h)\, \kappa_r = \Big(P^o_o = |{\vec P}_o|;
{\vec P}_o\Big)$ and $\epsilon^{\mu}_r(\vec h) = \Big(- h_r;
\delta^i_r - {{h^i\, h_r}\over {1 + \sqrt{1 + {\vec h}^2}}}\Big)$,
we get\medskip

$P^o_o = h^o\, \sqrt{{\vec \kappa}^2} + \vec h \cdot \vec \kappa$
\medskip

$P^i_o = h^i\, \Big[\sqrt{{\vec \kappa}^2} + {{\vec h \cdot \vec
\kappa}\over {1 + \sqrt{1 + {\vec h}^2}}}\Big] - \kappa_i$
\medskip

$K^{\mu}_o(P_o) = {1\over {2\, {\vec \kappa}^2}}\, \Big[h^{\mu}\,
\sqrt{{\vec \kappa}^2} + \epsilon^{\mu}_r(\vec h)\, \kappa_r\Big]$
\medskip

$K^o_o(P_o) = {1\over {2\, |h^o\, \sqrt{{\vec \kappa}^2} + \vec h
\cdot \vec \kappa|}}$\medskip

$K^i_o(P_o) = {1\over {2\, {\vec \kappa}^2}}\, \Big(h^i\,
\Big[\sqrt{{\vec \kappa}^2} - {{\vec h \cdot \vec \kappa}\over {1 +
\sqrt{1 + {\vec h}^2}}}\Big] + \kappa_i\Big)$\medskip

$\epsilon^{\mu}_{\lambda}(P_o) = A_{\lambda r}\,
\epsilon^{\mu}_r(\vec h)$ with $A_{\lambda r}\, \kappa_r = 0$ and
$\sum_r\, A_{\lambda r}\, A_{\lambda^{'} r} = \delta_{\lambda
\lambda^{'}}$\medskip

$\epsilon^{\mu}_{\lambda}(P_o) = \Big(\delta_{\lambda r} - {{P_o^r\,
P_{o \lambda}}\over {|{\vec P}_o|\, (|{\vec P}_o| + P^3_o)}} -
{{h^{\lambda}\, h_r}\over {1 + \sqrt{1 + {\vec h}^2}}} - {{P_{o
\lambda}\, h_r\, \sum_{a=1}^2\, P_{o a}\, h_a}\over {|{\vec P}_o|\,
(|{\vec P}_o| + P_o^3)\, (1 + \sqrt{1 + {\vec h}^2})}} + \delta_{3r}
{{P_{o \lambda}}\over {|{\vec P}_o|}} - {{h^3\, h_r\, P_{o
\lambda}}\over {(1 + \sqrt{1 + {\vec h}^2})\, |{\vec P}_o|}}\Big)\,
\epsilon^{\mu}_r(\vec h)$

\vfill\eject

\end{document}